\documentclass[prl,twocolumn,superscriptaddress]{revtex4-1}

\usepackage{graphicx,color}
\usepackage{amsmath}
\usepackage{amssymb}
\usepackage{natbib}

\newcommand{\HIO}{Ho$_2$Ir$_2$O$_7$}
\newcommand{\DIO}{Dy$_2$Ir$_2$O$_7$}
\newcommand{\DTO}{Dy$_2$Ti$_2$O$_7$}

\begin{document}

\title{Fragmented monopole crystal, dimer entropy and Coulomb interactions in Dy$_2$Ir$_2$O$_7$}

\author{V. Cathelin}
\affiliation{Institut N\'eel, CNRS \& Univ. Grenoble Alpes, 38042 Grenoble, France}
\author{E. Lefran\c cois}
\affiliation{Institut N\'eel, CNRS \& Univ. Grenoble Alpes, 38042 Grenoble, France}
\affiliation{Institut Laue Langevin, CS 20156, 38042 Grenoble, France}
\author{J. Robert}
\affiliation{Institut N\'eel, CNRS \& Univ. Grenoble Alpes, 38042 Grenoble, France}
\author{P. C. Guruciaga}
\affiliation{Centro At\'omico Bariloche, Comisi\'on Nacional de Energ\'{\i}a At\'omica (CNEA), Consejo Nacional de Investigaciones Cient\'{\i}ficas y T\'ecnicas (CONICET), Av.~E.~Bustillo 9500, R8402AGP San Carlos de Bariloche, R\'{\i}o Negro, Argentina}
\author{C. Paulsen}
\affiliation{Institut N\'eel, CNRS \& Univ. Grenoble Alpes, 38042 Grenoble, France}
\author{D. Prabhakaran}
\affiliation{Clarendon Laboratory, Physics Department, Oxford University,
Oxford, OX1 3PU, United Kingdom}
\author{P. Lejay}
\affiliation{Institut N\'eel, CNRS \& Univ. Grenoble Alpes, 38042 Grenoble, France}
\author{F. Damay}
\affiliation{CEA, Centre de Saclay, /DSM/IRAMIS/Laboratoire L\'eon Brillouin, 91191 Gif-sur-Yvette, France}
\author{J. Ollivier}
\affiliation{Institut Laue Langevin, CS 20156, 38042 Grenoble, France}
\author{B. F\aa k}
\affiliation{Institut Laue Langevin, CS 20156, 38042 Grenoble, France}
\author{L. C. Chapon}
\affiliation{Diamond Light Source Ltd., Harwell Science and Innovation Campus, Didcot, United Kingdom}
\affiliation{Institut Laue Langevin, CS 20156, 38042 Grenoble, France}
\author{R. Ballou}
\affiliation{Institut N\'eel, CNRS \& Univ. Grenoble Alpes, 38042 Grenoble, France}
\author{V. Simonet}
\affiliation{Institut N\'eel, CNRS \& Univ. Grenoble Alpes, 38042 Grenoble, France}
\author{P. C. W. Holdsworth}
\affiliation{Universit\'e de Lyon, ENS de Lyon, Universit\'e Claude Bernard, CNRS, Laboratoire de Physique, F-69342 Lyon, France} 
\author{E. Lhotel}
\affiliation{Institut N\'eel, CNRS \& Univ. Grenoble Alpes, 38042 Grenoble, France}

\begin{abstract}
Neutron scattering, specific heat and magnetisation measurements on both powders and single crystals reveal that \DIO\ realizes the  fragmented monopole crystal state in which antiferromagnetic order and a Coulomb phase spin liquid co-inhabit. The measured residual entropy is that of a hard core dimer liquid, as predicted.
Inclusion of Coulomb interactions allows for a quantitative description of both the thermodynamic data and the magnetisation dynamics, with the energy scale given by deconfined defects in the emergent ionic crystal. 
Our data reveal low energy excitations, as well as a large distribution of energy barriers down to low temperatures, while
the magnetic response to an applied field suggests that domain wall pinning is important; results that call for further theoretical modelling.  
\end{abstract}

\maketitle

The fractionalisation of microscopic elements into collective objects of reduced dimension has been a key concept in condensed matter for several decades \cite{Anderson87}. 
In three dimensions the emergence, in frustrated pyrochlore magnets, of effective fields with $U(1)$ symmetry \cite{Isakov04} provides an important, geometrically driven and experimentally relevant source of fractionalisation \cite{Hermele04}. In particular, in spin ice materials \cite{Harris97} and models \cite{Hertog00,Castelnovo08}, the associated topological charge, dressed by real magnetic flux provides the magnetic monopole excitations \cite{Ryzhkin05,Castelnovo08} which have been much studied over the last decade. In this case, the magnetic moment configurations follow closely the emergent field theoretic picture and appear to fragment into two orthogonal fluids via a Helmholtz decomposition \cite{Brooks14}. The two components act independently and in the right conditions can even order independently, giving the possibility of a monopole charge crystal \cite{Borzi13, Brooks14, Guruciaga14,Raban19}, an antiferromagnetically ordered phase \cite{Lefrancois17, Petit16, Canals16, Paddison16} which coexists with a ferromagnetically correlated Coulomb phase \cite{Henley10}.  

Pyrochlore iridates $R_2$Ir$_2$O$_7$, where the rare earth $R$ and iridium form interpenetrating pyrochlore structures, are ideal materials to generate such physics on the magnetic rare-earth sublattice. In these compounds, the Ir$^{4+}$ sublattice orders magnetically into an ``all-in$-$all-out'' configuration (AIAO), with spins oriented along the local $\langle 111 \rangle$ directions \cite{Tomiyasu12, Sagayama13, Lefrancois15, Guo16}, at temperatures between 30 and 150 K \cite{Matsuhira11} (Pr$_2$Ir$_2$O$_7$ being an exception). As the $R-R$ interactions are generally in the Kelvin range, a good starting approximation is to treat this order as a staggered magnetic field which favours the same AIAO configurations for the rare earth spins \cite{Tomiyasu12, Lefrancois15}. Within the monopole picture, this corresponds to a staggered chemical potential \cite{Raban19} which reduces the point group symmetry of the monopole sites and opens the door to the stabilisation of the fragmented monopole crystal phase when $R-R$ interactions are ferromagnetic. 

In this Letter, we show that \DIO\ realises such a fragmented monopole crystal state at temperatures below around 1 Kelvin. 
We show that half the total moment of the Dy$^{3+}$ ions is devoted to each of the magnetic sectors while specific heat measurements expose the predicted residual entropy, which is that of a hard core dimer fluid on the diamond lattice \cite{Brooks14, Jaubert15,Nagle66}. We model the results, including Coulomb interactions between monopoles \cite{Castelnovo08}, finding good qualitative agreement with experiment, with our analysis highlighting the role of long range interactions for both static and dynamic measurements. However, our analysis also reveals the existence of low and high energy excitations that are not accounted for by simple models. Our results are compatible with previous experiments on \HIO\ \cite{Lefrancois17}, but go considerably beyond them in presenting quantitative measures of both the Coulomb phase and the magnetic, ionic crystal. 

Both polycrystalline and small single crystal ($ \sim 0.01~{\rm mm}^{3}$) samples were used \cite{supmat}.
Polycrystalline samples were characterised by neutron diffraction on the G4.1 (LLB) diffractometer down to 70 mK, and by inelastic neutron scattering down to 1.6 K on IN4 and IN6 (ILL) \cite{doi_IN6, supmat}. The latter measurements allowed us to refine the Dy$^{3+}$ crystal electric field, giving an Ising ground state doublet with a magnetic moment $m=9.85~\mu_{\rm B}$ \cite{supmat}. 
Magnetisation measurements were performed down to 2~K on Quantum Design (QD) MPMS and VSM SQUID magnetometers, and between 90 mK and 4 K on purpose-built SQUID magnetometers equipped with a miniature dilution refrigerator \cite{Paulsen01}. 
Specific heat measurements were performed  between 0.4 and 20 K with a $^3$He QD PPMS on the same single crystal (of mass $0.27$~mg) as the QD VSM SQUID measurements. The specific heat of a pellet of Eu$_2$Ir$_2$O$_7$ powder was measured as a reference non-magnetic rare-earth. For the very low temperature SQUID measurements, several single crystals were coaligned. 

\begin{figure}
\includegraphics[width=8cm]{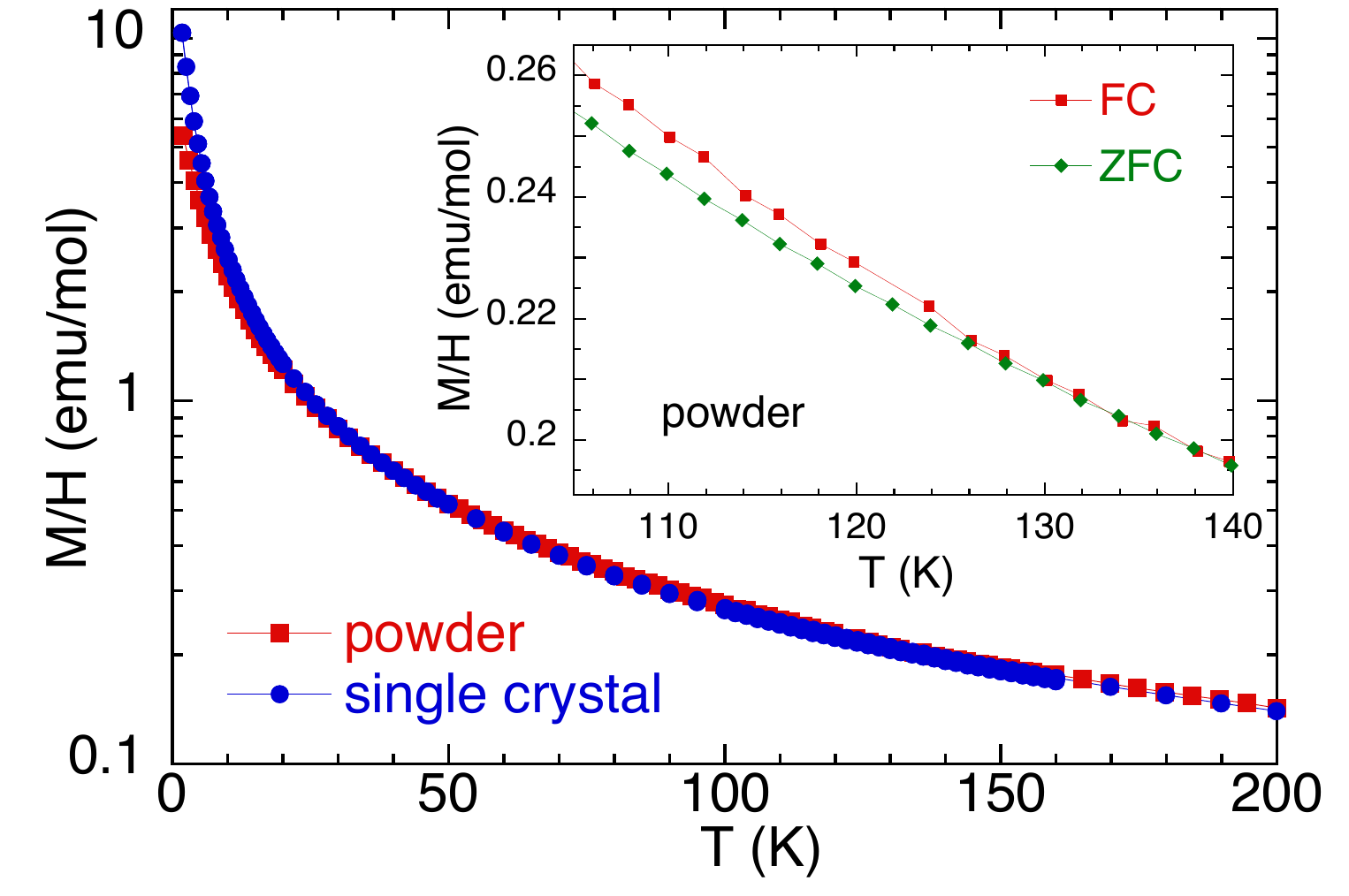}
\caption{\label{fig_MT} FC magnetisation $M/H$ vs temperature for the powder sample ($H=100$ Oe) and a single crystal ($H=1000$~Oe applied in an arbitrary direction) on a semi-logarithmic scale. Inset: zoom in the ZFC-FC magnetisation of the powder sample ($H=100$~ Oe). }
\end{figure}

Like in other pyrochlore iridates \cite{Matsuhira11}, the iridium AIAO ordering manifests through a small irreversibility in the  zero field cooled -- field cooled (ZFC-FC) magnetisation of the powder sample, below about 125 K (see Figure \ref{fig_MT}), a slightly smaller temperature than the 134~K reported in Ref. \onlinecite{Matsuhira11}.
The irreversibility is very small compared to this earlier study while no irreversibility could be detected for single crystals. The ZFC-FC irreversibility has been proposed to be due to structural defects and domain-walls, which modify the iridium molecular field felt by the rare earth ions, resulting in the enhancement of their polarization with decreasing temperature \cite{Yang17, Zhu14, Lefrancois15}. This scenario would suggest that our samples are cleaner than those used in previous reports.

\begin{figure}
\includegraphics[width=8cm]{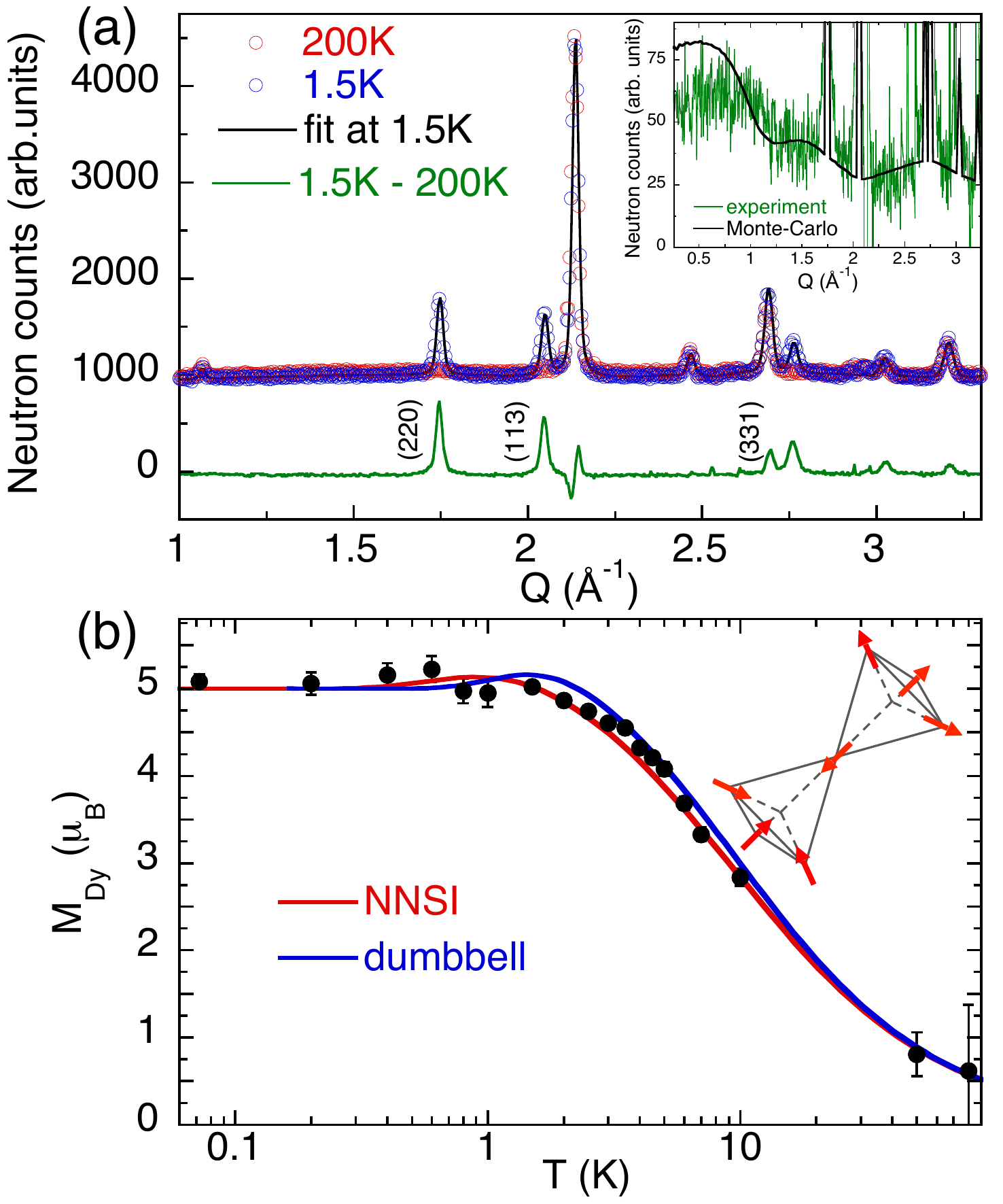}
\caption{\label{diff} (a) Diffractograms at $T=1.5$ (blue) and 200~K (red), and difference between 1.5  and 200 K (green). The black line is the refinement obtained at 1.5 K. Inset: Zoom of the difference, corrected from the paramagnetic scattering (green). The black line is the powder average magnetic scattering function from Monte Carlo calculations in the NNSI model for $T/J_{\rm eff} = 0.05$ with $h_{\rm loc}/J_{\rm eff} = 4.5$.
(b) Refined Dy$^{3+}$ ordered magnetic moment vs temperature between 80~K and 60~mK. Lines are the calculated ordered moment in the NNSI model for ${\cal J}_{\rm eff}=1.1$ K and $h_{\rm loc}/{\cal J}_{\rm eff}$= 4.5 (red) and in the dumbbell model for $\mu=-4.40$ K and $\Delta=4.95$ K (blue). Inset: AIAO configuration on two tetrahedra.}
\end{figure}

Magnetic Bragg peaks appear in powder neutron diffraction measurements below about 100 K (see Figure \ref{diff}(a)). {\sc Fullprof} refinements \cite{Rodriguez-Carvajal93} with a ${\bf k}={\bf 0}$ propagation vector give an AIAO magnetic structure  \cite{Lefrancois15, Guo16, Yan17} (shown in the inset of Figure \ref{diff}(b)) for both the Dy and Ir ions. The low temperature iridium ordered moment is found to be constant in the analysis range ($T<80$~K) and equal to $m_{\rm Ir}=0.34\pm0.14~\mu_{\rm B}$. The temperature dependence of the ordered moment per Dy$^{3+}$ ion $m_{\rm Dy}$ between 10 and 80 K (Figure \ref{diff}(b)) is characteristic of field induced order \cite{Lefrancois15}. At lower temperature, Dy-Dy interactions favor a spin ice state, which competes with this field induced state, leading to a saturation of $m_{\rm Dy}$ below $T=1.5$ K to the value of $5 \pm 0.1~\mu_{\rm B}$, that is, to half of the total moment, as expected in the fragmentation scenario. Some diffuse magnetic signal persists down to the lowest temperature (see inset of Figure \ref{diff}(a)). These measurements thus provide two essential fingerprints for the stabilization of a fragmented crystal  state in \DIO: AIAO ordering accounting for half of the magnetic moment coexisting with a correlated spin liquid phase. 

\begin{figure}
\includegraphics[width=8cm]{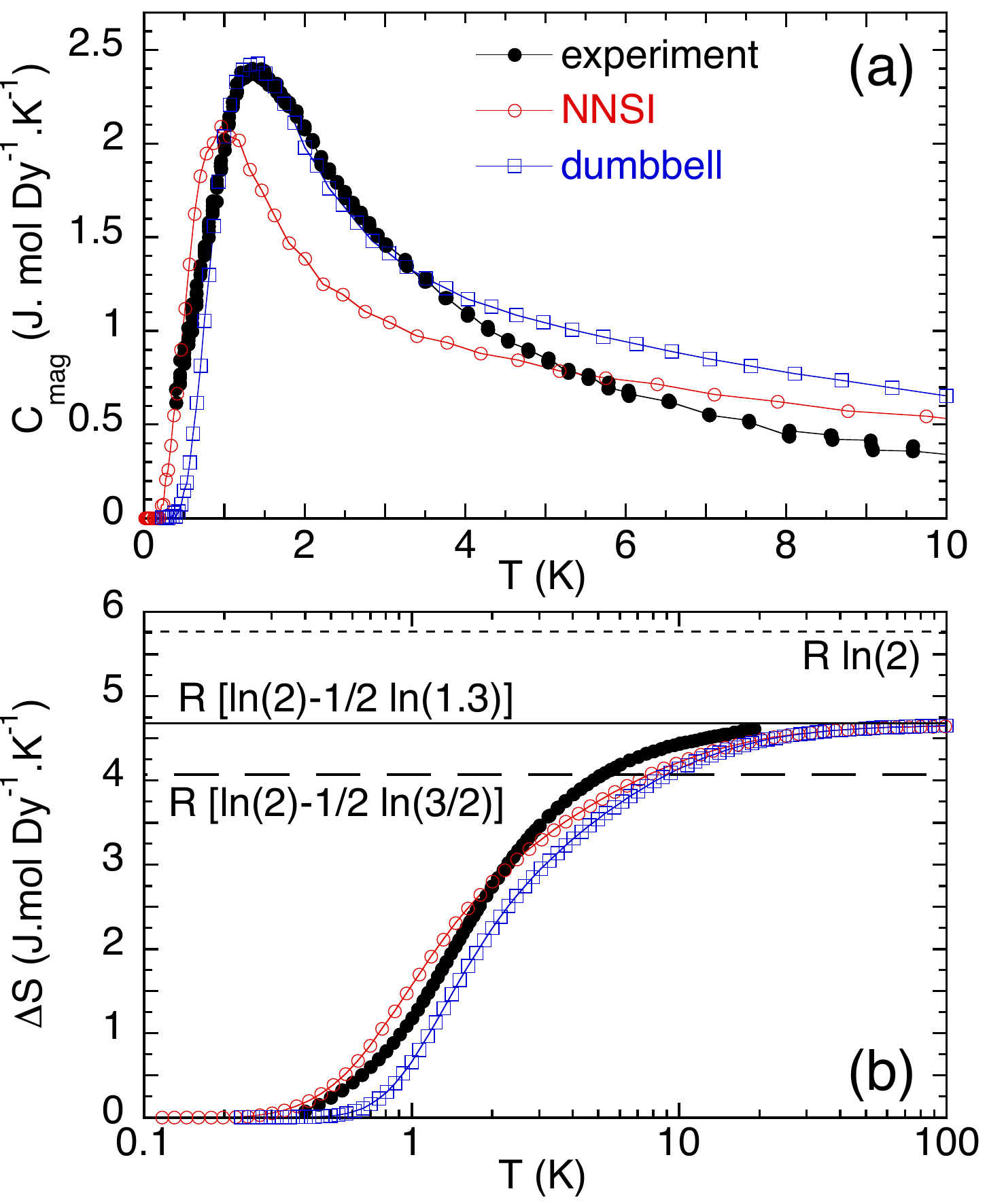}
\caption{\label{fig_C} (a) $C_{\rm mag}$ vs $T$. The experiment (black dots) is compared to the NNSI (red circles, $h_{\rm loc}/{\cal J_{\rm eff}}=4.5$ and ${\cal J_{\rm eff}}=1.1$ K) and the dumbbell (blue squares, $\mu=-4.40$~K and $\Delta =4.95$ K) models. Specific heat data of Eu$_2$Ir$_2$O$_7$ with non magnetic Eu were subtracted from the original data to extract the Dy magnetic contribution. (b) Entropy obtained from the integration of the above curves (semilogarithmic scale). $R \ln(2)$ corresponds to the full spin entropy, $R [\ln(2) - 1/2 \ln(1.3)]$ to the fragmented entropy and $R [\ln(2) - 1/2 \ln(3/2)]$ to the Pauling entropy of ice.}
\end{figure}

As the local field lowers the symmetry of the monopole sites to that of the zinc-blende structure \cite{Sands69}, a thermal phase transition is not required and none is observed in specific heat measurements (see Figure \ref{fig_C}(a)).
However, as one enters fully into the fragmented phase a broad peak is observed, with a maximum at about $T=1.4$ K. This is reminiscent of the signal observed in classical spin ice \cite{Ramirez99} but is even broader, spreading out to much higher temperature, reflecting the energy scale of the local field.

Our low temperature results differ from previous studies, which report a broad maximum at about 5~K in the susceptibility \cite{Matsuhira11b} or a sharp peak in the specific heat at 1.2 K \cite{Yanagishima01}. Nevertheless, our measurements performed on both a powder and single crystals synthesized in different laboratories are consistent with each other. In addition, most of our observations can be accounted for by the model developed below.  

We model the magnetic Dy-Ir interaction by a temperature independent mean field term. This is most easily considered using the nearest neighbour spin ice model (NNSI) \cite{Lefrancois17}:
\begin{equation} 
\label{NNSI}
{\cal H}={\cal J}_{\rm eff} \sum_{<i,j>}{\sigma_i  \sigma_j } - h_{\rm loc} \sum_{i}{\sigma_i}
 \end{equation}
 where ${\cal J}_{\rm eff}$ is an effective, ferromagnetic nearest neighbor coupling,
${\sigma}_i=1$ ($-1$) is a reduced spin variable pointing in (out) of an up tetrahedron \cite{Jaubert11} and  $h_{\rm loc}$ is a staggered magnetic field coming from the iridium ions. However, the monopole approximation for spin ice, including long range interactions is captured by 
the dumbbell model \cite{Castelnovo08, Kaiser18, Castelnovo19}. Here magnetic charge $Q_i$
 sits at the vertices $i$ of the diamond lattice, dual to the pyrochlore lattice and the spin Hamiltonian is replaced by 
\begin{equation} 
\label{dumbbell}
{\cal H}_{\rm db}=\frac{u}{2} \sum_{i \neq j} \left( \frac{a}{r_{ij}} \right) \hat{n}_i \hat{n}_j - \mu \sum_i \hat{n}_i^2 - \Delta \sum_{i=1,N_0} (-1)^i \hat{n_i} ,
\end{equation} 
where $\hat{n}_i=Q_i/Q=0,\pm 1, \pm 2$ is a site occupation variable, $Q=2m/a$ the monopole charge, $u=\frac{\mu_0 Q^2}{4\pi a}=2.82$~K the Coulomb energy scale, $\mu<0$ the chemical potential \cite{Raban19} and $N_0$ the number of tetrahedra.  The staggered chemical potential $\Delta$ replaces $h_{\rm loc}$ giving  an energy difference for monopole creation on the two sublattices of the diamond lattice \cite{Brooks14}. 
Note that, although the field acts on a dipole and chemical potential on a monopole, when reduced to units of energy they are equal: $h_{\rm loc}=\Delta$ \cite{Raban19}.

We have fitted the experimental results for the Dy ordered moment with data from the NNSI, with 
${\cal J}_{\rm eff}=1.1 \pm 0.1$ K and $h_{\rm loc}=4.95 \pm 0.25$ K. For the dumbbell model, parameters were chosen to simultaneous reproduce both the magnetisation and the specific heat, giving  $\mu=-4.40 \pm 0.10$ K and $\Delta=4.95 \pm 0.15$~K (see Figure \ref{diff}(b) and Figure \ref{fig_C}(a)).
The values of ${\cal J}_{\rm eff}$ and $\mu$ are close to the estimates for \DTO\ \cite{Melko04}. The $h_{\rm loc}/{\cal J}_{\rm eff}$ ratio is the same as for \HIO\ and these values place \DIO\ deep in the predicted fragmented crystal phase at low temperature \cite{Lefrancois17,Raban19}. 

The NNSI model fits the temperature dependence of $m_{\rm Dy}$ quite accurately but in doing so 
gives a poor representation of the specific heat (see Figure \ref{fig_C}(a)), as was the case for \DTO\ \cite{Melko04}. Introducing long range interactions, the dumbbell model reproduces both  $m_{\rm Dy}$ and the specific heat peak height and position, although the agreement is less convincing in the wings at high and low temperature. Above 4 K the model specific heat remains considerably higher than that of the experiment, indicating that correlations exist out to even higher temperatures.
More surprisingly, while the model specific heats drop exponentially at small temperature, the experimental data appears to fall more slowly, retaining entropy down to lower temperatures. This indicates that low energy excitations are present, which are not accounted for theoretically. These may originate from corrections to the dumbbell model which lift the degeneracy of the Coulomb phase,
to structural defects, or to low energy excitations in the iridium sector that are not accounted for.  

In the monopole crystal phase it is predicted that the closed loops of virtual spin flips should induce a residual entropy equal to that of an ensemble of hard core dimers on a diamond lattice, $S\approx \frac{1}{2} \ln(1.3) =0.131$ per spin \cite{Brooks14, Jaubert15,Nagle66} and the models have this ground state entropy built into them. This is confirmed in Figure \ref{fig_C}(b) where we show  the entropy recovered 
through integrating $\frac{C}{T}$ for both experiment and simulation.
Experimentally, despite the apparent quantitative difference with the models we also recover this residual entropy to an excellent approximation.

\begin{figure}
\includegraphics[width=8cm]{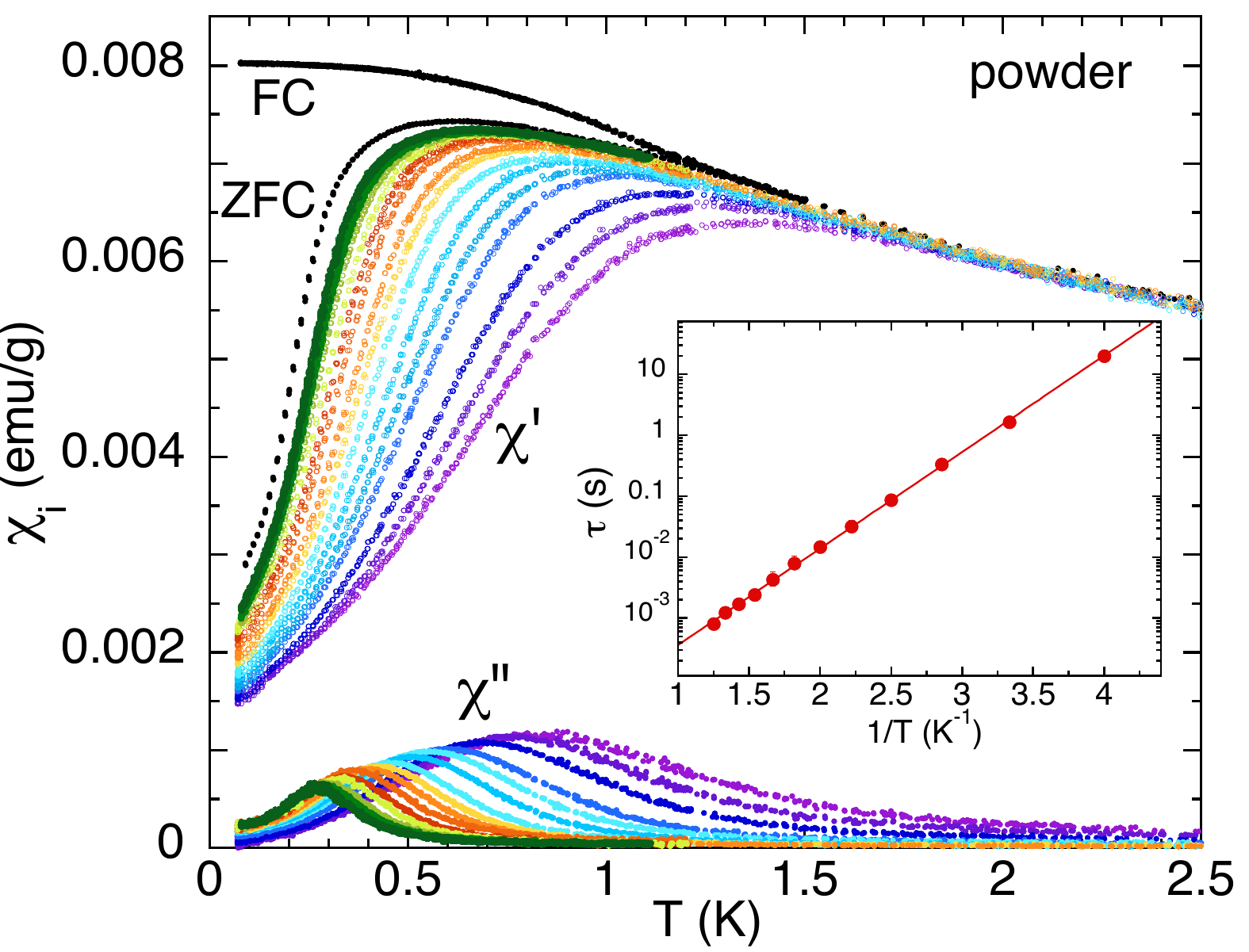}
\caption{\label{fig_Xac}  ZFC-FC dc susceptibility (black points, $H=50$~Oe) and ac susceptibility, $\chi'$ and $\chi''$ (coloured symbols, frequencies $f$ between 0.0057 and 211 Hz, $H_{\rm ac}=1$ Oe) vs temperature for the powder sample. Data were corrected for demagnetisation effects with a demagnetisation factor $N=0.1$ cgs. Inset: Relaxation time $\tau=1/2 \pi f$ vs $1/T$, obtained from the maximum of $\chi''$ vs $f$ measurements at fixed temperature \cite{supmat}. The red line is a fit to the Arrhenius law $\tau=\tau_0 \exp{(E/T)}$ with $\tau_0=9.4\times 10^{-6}$ s and $E=3.6$ K. }
\end{figure}

The dynamics of the fragmented state can be probed with magnetisation and ac susceptibility measurements. A freezing is observed when the system enters the fragmented crystal state, which manifests as a separation at $T=1.4$ K between the ZFC and FC magnetisations measured when cooling from 4 K (see Figure~\ref{fig_Xac}). 
Although the shape of the curve is slightly different for powders and single crystals, the ratio $M_{\rm ZFC}/M_{\rm FC}$ is the same in both cases and  reaches about 0.2 at 80~mK \cite{supmat}. 
The ZFC magnetisation remains finite down to 80 mK, contrary to Dy$_2$Ti$_2$O$_7$, where it falls to zero below 300 mK \cite{Snyder04}.  This indicates that additional degrees of freedom exist that help magnetisation to relax, consistently with our observations for the specific heat. 

The ac susceptibility, shown in Figure \ref{fig_Xac}  exhibits a frequency dependence that can be accurately described by a thermally activated process, above an energy barrier $E=3.6$ K. 
This dynamics can be understood through the propagation of magnetically charged, deconfined defects in the monopole crystal \cite{Jaubert15}. In the dumbbell model, the lowest energy excitation is a double monopole with energy \cite{supmat} 
\begin{equation}
E_{\rm db}=-(3\mu + \Delta) - u \alpha, \; \alpha=1.638,
\end{equation} 
which gives $E_{\rm db}=3.63$ K, in remarkable agreement with experiment. However, a word of caution is required; the propagation of the excitation, through a single spin flip, creates a hole of energy $E^{'}_{\rm db}=\mu+\Delta+u\alpha=5.2$ K. In order to avoid this higher energy scale the dynamics would have to involve double spin flips \cite{Hermele04,Jaubert15,supmat}. The NNSI model underestimates these energies giving $E_{\rm NNSI}= 1.65$~K and $2.75$~K respectively \cite{Lefrancois17},  illustrating the importance of the Coulomb interaction between the magnetic charges. A Cole-Cole analysis of the ac susceptibility data shows that a large distribution of time scales exists, which broadens as the temperature decreases \cite{Cole41, Dekker89, supmat}.  Just as for spin ice, this is compatible with quasiparticle hopping via a range of microscopic time scales \cite{Jaubert09,Bovo13,Dusad19}.  

\begin{figure}
\includegraphics[width=8cm]{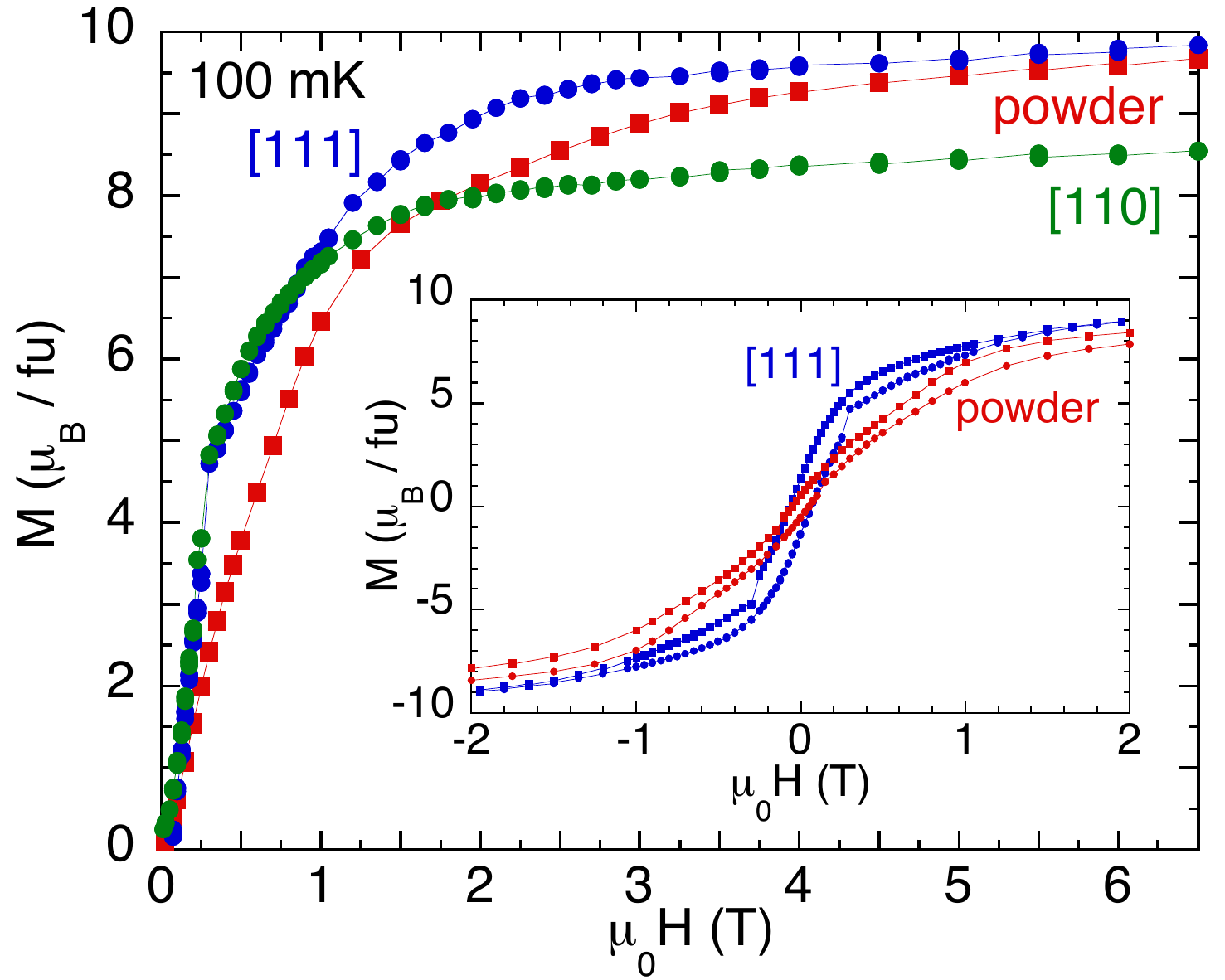}
\caption{\label{fig_MH} Magnetisation vs field at $T=100$ mK for the powder sample and the single crystal (${\bf H} \parallel [110]$ and $[111]$) measured starting from a ZFC state, except for the $[111]$ direction where measurements were made from saturation in a negative field.  
Inset: Zoom in the hysteresis loop for the powder and ${\bf H} \parallel [111]$. Curves have been symmetrized. }
\end{figure}

The phase diagram of the fragmented crystal, as a function of applied magnetic field ${\bf H}$, is expected to be rich \cite{Lefrancois17,Brooks14,supmat}. The magnetic field couples independently to the two fragmentation sectors, 
remarkably providing a staggered chemical potential for the monopoles \cite{Castelnovo08,Raban19} in competition with the staggered internal field. As a consequence, a field placed along 
$[111]$ (forward) and $-[111]$ (reverse) directions are inequivalent, working with or against the internal field. In the forward direction the monopole configuration is unchanged by the field so that the magnetisation should saturate via a Kasteleyn transition \cite{Brooks14,Moessner03} at low field. For $T=0.1$ K the saturation field is only 6 mT \cite{supmat}.  
In the reverse direction the field generates a reduced effective $\Delta$, forcing the system back into the spin ice phase above a first threshold and into a monopole crystal going against the staggered field above a second threshold.
This reorganisation leads to three magnetisation plateaux \cite{Lefrancois17} and, for long range interactions further phase transitions  \cite{Raban19}. We predict a first plateau with $M=m/6$ for low field, jumping to a second at $m/3$ for $\mu_0H\approx 1.3$ Tesla and to a third at saturation, for $\mu_0H\approx 3.2$ Tesla \cite{supmat}.  

The availability of single crystals allows us to test these predictions. Data for fields placed along the $[111]$ and $[110]$ directions, together with measurements from powder samples are shown in Figure \ref{fig_MH}. The saturated magnetisation per Dy approaches the expected values, $M_{[111], {\rm pwd}}=m/2$ and $M_{[110]}=m/\sqrt{6}$ \cite{Harris98} for fields above $3$ T.
The predicted magnetisation plateaux are not observed, although the initial slope is steep, and in the powder data a change of slope is observed at around $1.5$~T, corresponding to the centre of the second plateau. 

An explanation for the absence of plateaux could be the presence of a partially frozen mosaic of ``AIAO / AOAI" iridium domains (as observed in Nd$_2$Ir$2$O$_7$ \cite{Ma15}) which drive domains of monopole crystal order. In this case, the two kinds of domains would see the applied field as a forward or a reverse field. If completely frozen, the reverse domains would dominate the field response, resulting in the observation of plateaux for arbitrary magnetisation values. Partial reorganisation of the domain structure would result in field induced evolution of the fraction of the sample following the forward response scenario,  masking the plateaux of the reverse response.  
Such a mixed response would terminate for fields around the upper threshold of $3.2$ T, which is consistent with the experimental results. 

We observe a narrow hysteresis on field sweeping (see inset of Figure \ref{fig_MH}) which is consistent with the partial pinning of domains.
It is accompanied in single crystals by small magnetisation ``avalanches'', driven by self heating as for \DTO\  \cite{Slobinsky10,Jackson14}, although the effect is less dramatic here, possibly due to the large thermal conductivity of the iridates or to additional relaxation channels offered by corrections to the simple models.   
The pinning appears stronger in the powder, where the remanence of the plateau is observed, which is consistent with our results at the iridium transition - see Figure \ref{fig_MT}. 

In conclusion, \DIO\ stabilises the fragmented monopole crystal state. Our analysis shows that both static and dynamic properties within this phase are governed by long range interactions, captured in a first approximation by the monopole picture of spin ice.
However, our measurements show evidence of low energy excitations which are not generated by the model. Magnetisation curves measured on single crystals do not show evidence of predicted magnetisation plateaux, or of the reduced point group symmetry of the monopole crystal. This suggests that the role of the iridium has to be examined further, both at the microscopic level and in terms of its domain structure and dynamics.    

\acknowledgments
{We acknowledge A. Hadj-Azzem for his help in the compound synthesis, J. Debray for the orientation of the single crystals, P. Lachkar for his help with the PPMS, and F. Museur for discussions. V. Cathelin, P.C.W. Holdsworth, E. Lhotel and C. Paulsen acknowledge financial support from ANR, France, Grant No. ANR-15-CE30-0004. P.C.W. Holdsworth, P. Lejay, E. Lhotel, and P.C.W. Holdsworth acknowledge financial support from ANR, France, Grant No. ANR-19-CE30-0040. D. Prabhakaran acknowledges financial support from EPSRC, UK, Grant No. EP/N034872/1. }



\clearpage

\renewcommand{\thefigure}{S\arabic{figure}} 
\renewcommand{\theequation}{S\arabic{equation}} 

\makeatletter
\renewcommand{\citenumfont}[1]{S#1}
\renewcommand{\@biblabel}[1]{\quad S#1. }
\makeatother

\setcounter{figure}{0} 
\setcounter{equation}{0} 

\onecolumngrid
\begin{center} {\bf \large Fragmented monopole crystal, dimer entropy and Coulomb interactions in Dy$_2$Ir$_2$O$_7$ \\ \medskip Supplementary Material} \end{center}
\vspace{0.5cm}
\twocolumngrid

\section{Synthesis}
Polycrystalline samples of \DIO\ were synthesized at the Institut N\'eel by a mineralization process, following the procedure described in Ref. \onlinecite{suppLefrancois17}. The lattice parameter and the $x$ coordinate of the 48f oxygen atom were found to be $a=10.181$ \AA\ and $x=0.335$ at $T=1.5$~K.\\

Small single crystals ($ \sim 0.01~{\rm mm}^{3}$) were synthesized at Oxford starting from a polycrystalline \DIO\ powder sample prepared in the stoichiometric ratio of 1:1.05 using high purity ($>99.99~\%$) Dy$_2$O$_3$ and IrO$_2$ chemicals. The powders were then thoroughly mixed along with 0.1~g of KF (for 5 g) inside an Argon glove box and pressed into 15 mm diameter pellets. Using a Pt crucible the pellets were sintered at 1100 $^{\circ}$C for 100 h.  The single phase pyrochlore powder was used as a starting material and mixed with KF flux in the ratio 1:200 and packed into a Pt crucible with a tightly fitted lid \cite{Millican07}.  The crucible was placed inside a chamber furnace and heated to 1050 $^{\circ}$C and after holding for 10 h, it was cooled down to 875 $^{\circ}$C at 1 $^{\circ}$C/h rate and finally to room temperature at 60 $^{\circ}$C/h. Octahedral shaped single crystals were separated from the flux using hot water.  Phase purity of the powder and single crystal sample was characterised using PANalytical and Supernova x-ray diffractometers respectively. 

\begin{figure}[b]
\includegraphics[width=7.5cm]{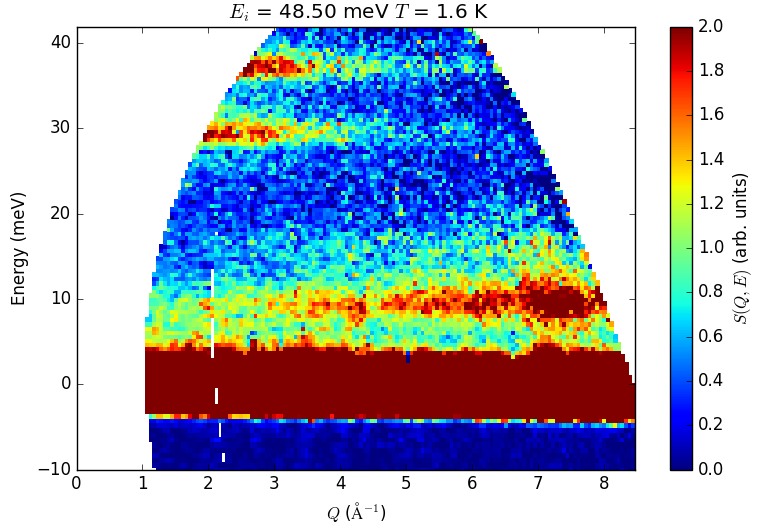}
\caption{Inelastic neutron scattering measurements of $S(Q,E)$ measured on IN4 at $T=1.6$ K with an incoming wavelength of 1.3 \AA.}
\label{INS}
\end{figure}

\section{Neutron diffraction measurements}
Neutron powder diffraction (NPD) measurements were carried out on the G4.1 diffractometer (LLB-Orph\'ee, France), equipped with an orange cryostat (for experiments in the 1.5 - 300 K range) or a Cryoconcept-France HD dilution refrigerator (100 $\mu$W@100 mK) for experiments in the 72 mK - 4 K range. For the experiments in the orange cryostat, the sample was put in a 3 cm diameter sample holder. For the experiments inside the dilution fridge, the sample was put in a specific 1 cm diameter vanadium cell, in 14 bars of He gas at ambient temperature, to ensure proper thermalisation. The working wavelength of G4.1 was 2.427 \AA. Rietveld refinements of the powder diffractograms were performed with the {\sc Fullprof} suite \cite{Carvajal93}.

\section{Inelastic neutron scattering measurements and Crystal Electric Field}
Crystal Electric Field (CEF) parameters were refined following the procedure detailed in the Supplementary note 2 of Ref. \onlinecite{suppLefrancois17}.

Inelastic neutron scattering measurements were performed down to 1.6 K on IN4 and IN6 (ILL) with incident wavelengths $\lambda_i=0.74, 0.9, 1.3, 1.8$ \AA\ and $\lambda_i=5.1$~\AA\ respectively \cite{suppdoi_IN6}. Two clear CEF modes are observed at 29.5 and 37 meV (See Figure \ref*{INS}). No dispersion of the crystal field excitations were visible within the resolutions of the inelastic neutron scattering experiments.

The CEF Hamiltonian for f electrons in the D$_{3d} (\bar{3}m)$ point group symmetry of the 16d Wyckoff site occupied by the Dy$^{3+}$ ions in the Dy$_2$Ir$_2$O$_7$ crystal writes:
\begin{equation}
\begin{aligned}
\mathcal{H}_{\rm CEF}  = & B_2^0\mathbf{C}_2^0 +  \\
     &  B_4^0\mathbf{C}_4^0 + B_4^3(\mathbf{C}_4^3 - \mathbf{C}_4^{-3}) +  \\
     & B_6^0\mathbf{C}_6^0 +  B_6^3(\mathbf{C}_6^3 - \mathbf{C}_6^{-3}) + B_6^6(\mathbf{C}_6^6 - \mathbf{C}_6^{-6})
\end{aligned}
\end{equation}
when the quantization axis is chosen along the local 3-fold axis. The $\mathbf{C}_k^q$
stand for Wybourne operators that, in a spatial rotation, transform like the spherical harmonics  $Y_k^q$. The $B_k^q$ are the (real) CEF parameters.

\begin{figure*}[t!]
\includegraphics[height=5cm]{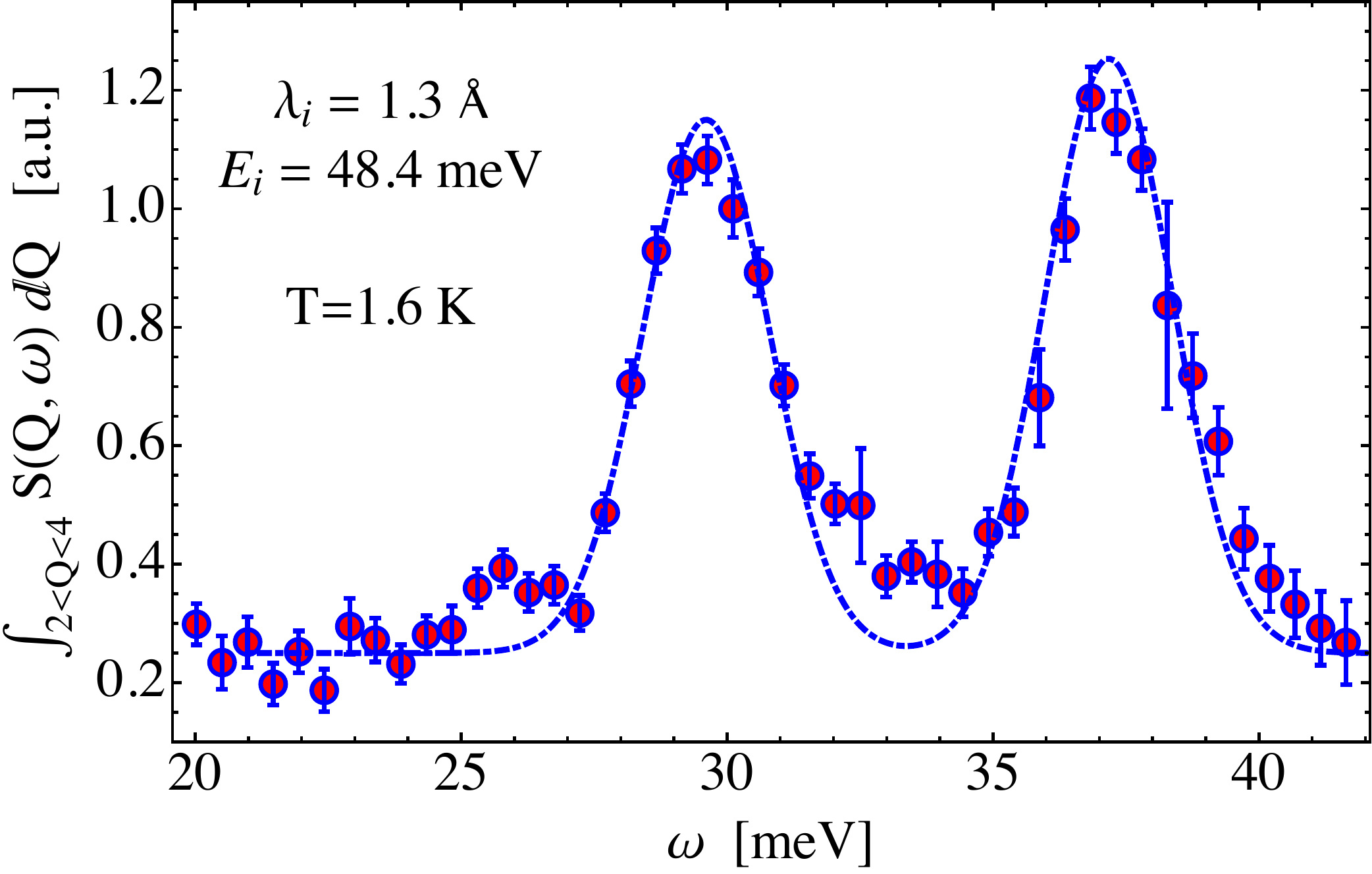} \qquad 
\includegraphics[height=5cm]{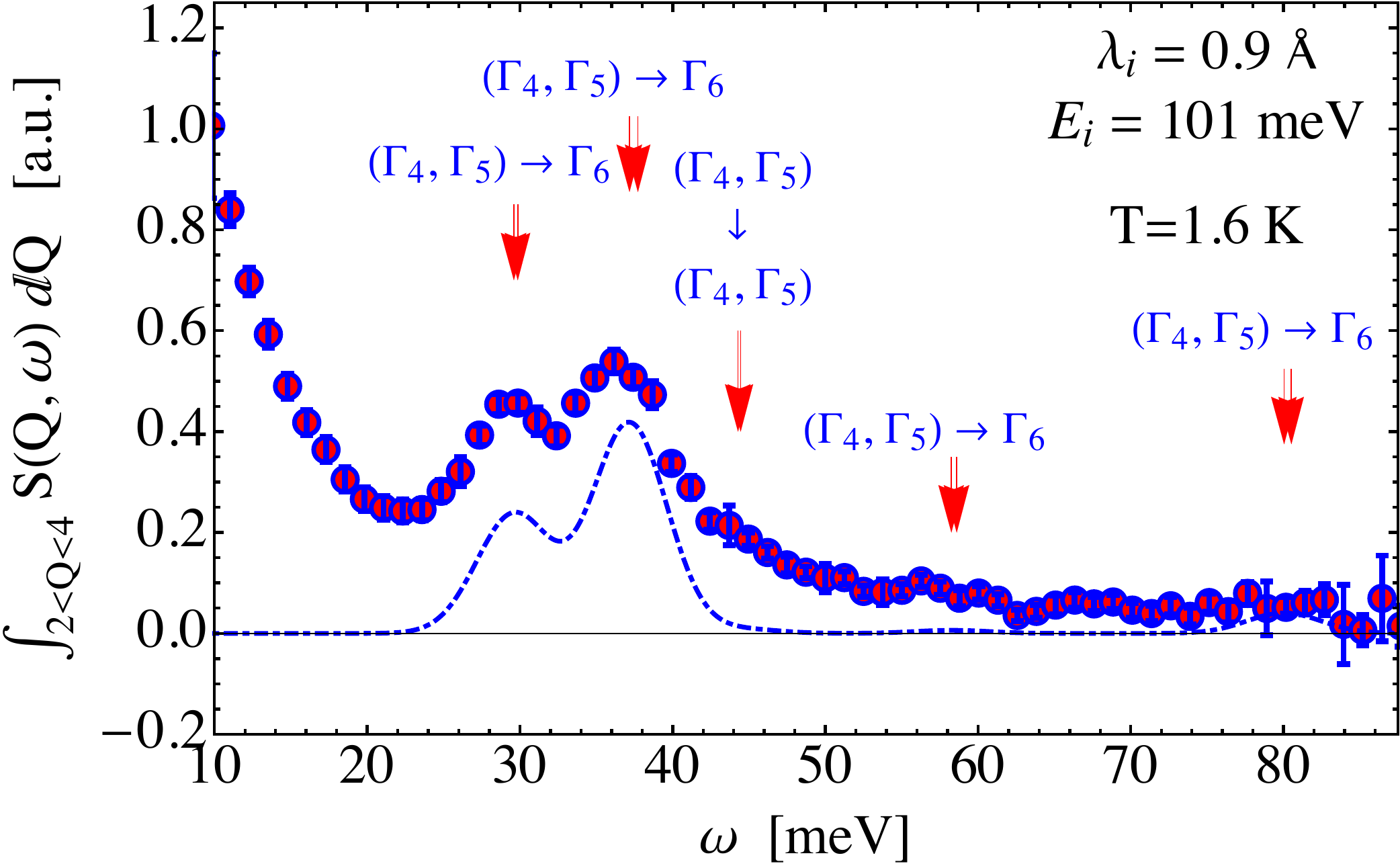}
\caption{Integrated inelastic neutron scattering intensity $\int{S(Q,\omega)dQ}$ measured on IN4 at $T=1.6$ K (red points) with incoming wavelengths $\lambda_i=1.3$ \AA\ (left) and 0.9 \AA\ (right). The data were integrated in the $Q=2-4$ \AA$^{-1}$ range. 
The dashed lines stand for the integrated neutron scattering function associated with the CEF transitions using the CEF parameters given in the text. The symbols $\Gamma_i$ ($i=4,5,6$) stand for the irreducible representations discussed in the text. }
\label{CEF_fit}
\end{figure*} 

$\mathcal{H}_{\rm CEF}$ lifts the degeneracy of the ground state multiplet $^6H_{15/2}$ ($S=5/2, L=5, J=15/2$) of the ion Dy$^{3+}$ into three doublets $(\Gamma_4, \Gamma_5)$ and five doublets $\Gamma_6$. $\Gamma_4$ and $\Gamma_5$ are the even-parity one-dimensional irreducible representations of the double point group generated from D$_{3d} (\bar{3}m)$ joined together by time inversion to form a corepresentation $(\Gamma_4, \Gamma_5)$ and $\Gamma_6$ is the even-parity two-dimensional irreducible representation of the same double group forming a corepresentation by its own. 

The $B_k^q$ CEF parameters were numerically estimated by restricting to the $^6H_{15/2}$ through reverse Monte Carlo from the energy and intensity of the excitations detected by neutron.
The best fits are obtained for: \\ $B_2^0 = 64.1 \pm 0.4$~meV, $B_4^0 = 307.0 \pm 1.5$~meV, $B_4^3= 90.3 \pm 3.0$~meV, $B_6^0 = 129.4 \pm 0.6 $~meV, $B_6^3 = -90.7 \pm 3.0$~meV, $B_6^6 = 69.5 \pm 1.5 $~meV (See Figure \ref*{CEF_fit}). 

This leads to the ground state $(\Gamma_{4}, \Gamma_{5})$ doublet :
\begin{align*}
| \pm \rangle \approx &\pm 0.98 | \pm 15/2 \rangle \mp 0.18 | \pm 9/2 \rangle \mp 0.02 | \pm 3/2 \rangle\\
 & \pm 0.03 | \mp 3/2 \rangle
\end{align*}
to which a pseudo-Ising magnetic moment of magnitude $9.85~\mu_{\rm B}$ is associated, aligned along the local 3 fold axes. The parallel and perpendicular Land\'e factors are calculated to $g_{\parallel} = 19.71$ and $g_{\perp} \approx 0$.

\section{Single crystal low temperature magnetization and susceptibility}
\begin{figure}[h!]
\includegraphics[width=7.5cm]{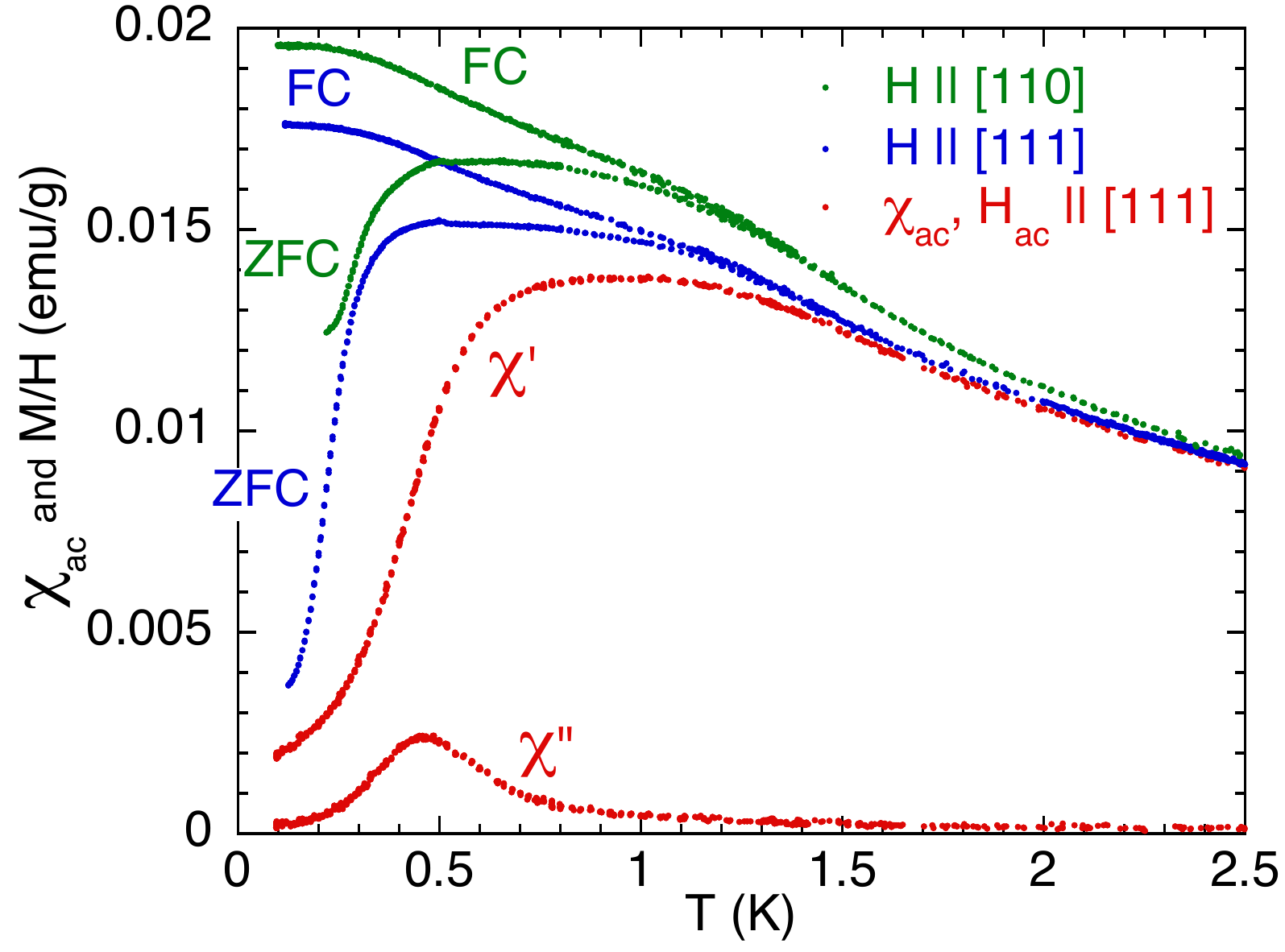}
\caption{ZFC-FC magnetization vs temperature for the single crystals measured with $H_{\rm dc}=100$ Oe applied along the $[110]$ (green) and $[111]$ (blue) directions, together with the ac susceptibility (red) at $f=2.11$ Hz with $H_{\rm ac}=1.82$ Oe along the $[111]$ direction.  (The starting ZFC magnetization measured along the $[110]$ direction is larger because the measurement only started at 225 mK). 
 }
\label{XT}
\end{figure}

The ZFC-FC magnetization for the single crystal was measured with the field applied along the $[111]$ and $[110]$ directions (see Figure \ref*{XT}). In both directions, the ZFC-FC separation is observed at the same temperature as the powder sample. 

\begin{figure}[t]
\includegraphics[width=7.5cm]{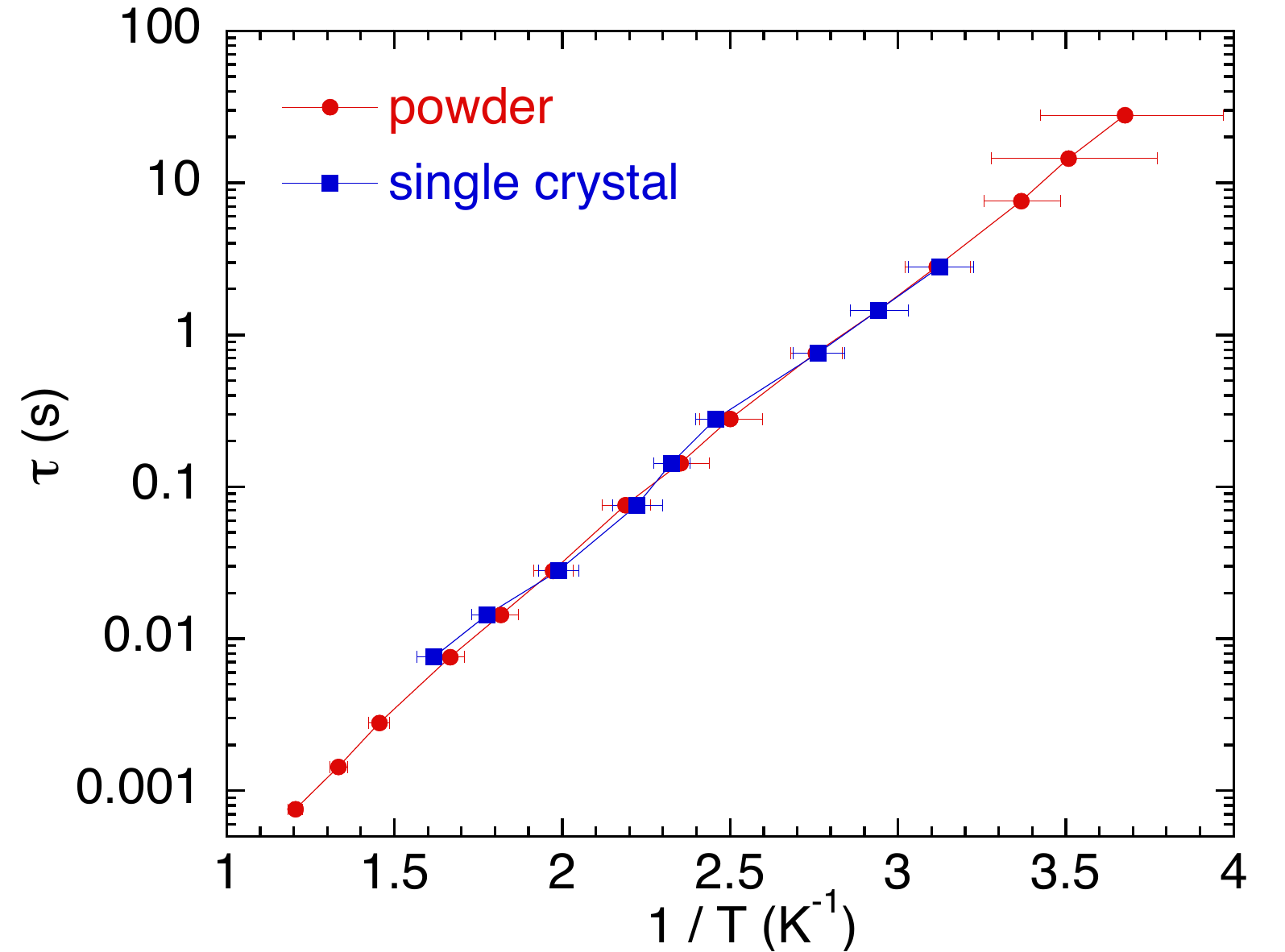}
\caption{$\tau=1/2 \pi f$ vs $T$ obtained from the maximum of $\chi''$ in the susceptibility vs temperature curves for the powder (red dots) and the single crystal (blue squares). 
 }
\label{tauvsT}
\end{figure}
\begin{figure}[t!]
\includegraphics[width=7.5cm]{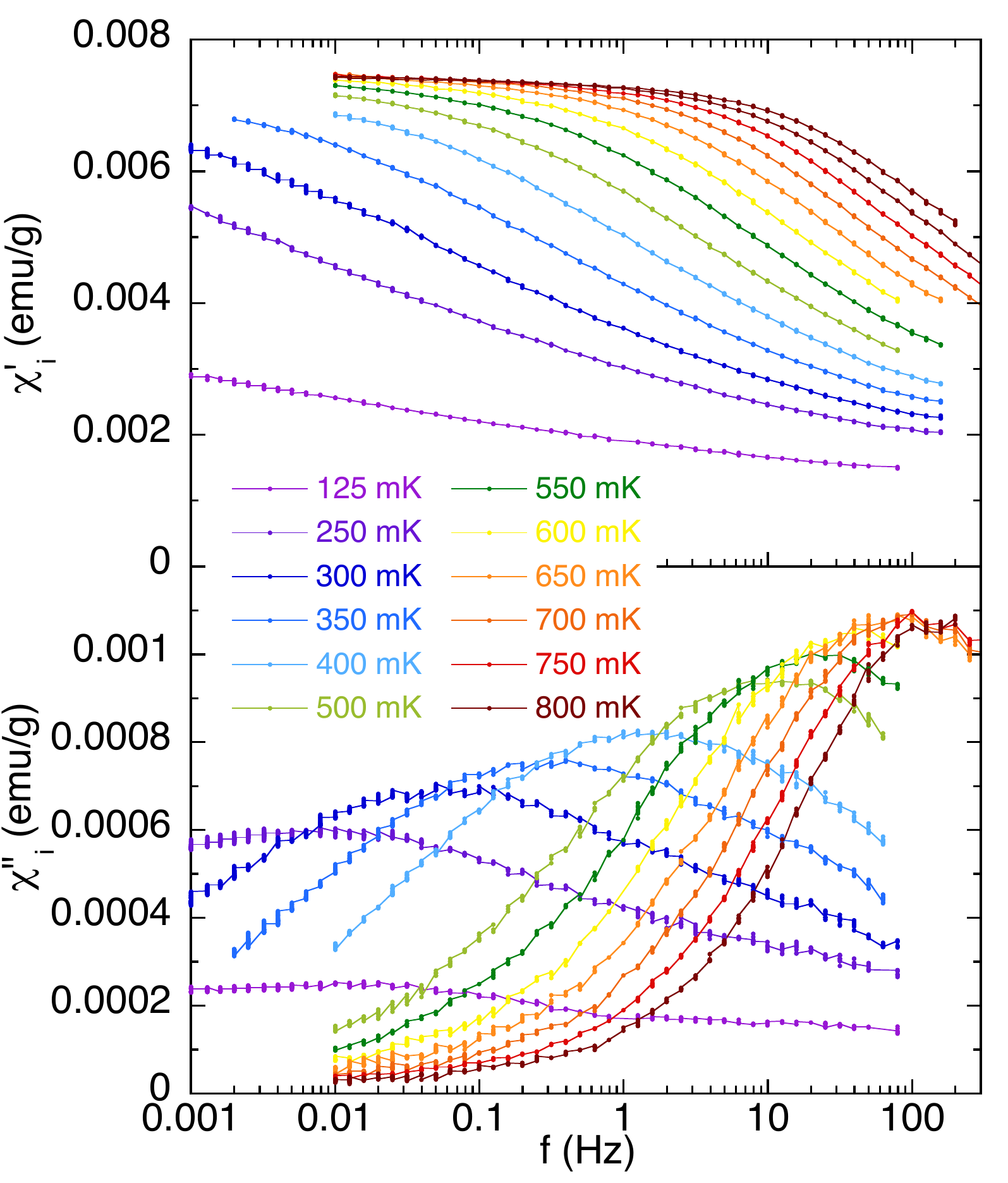}
\caption{$\chi'$ (top) and $\chi''$ (bottom) vs $f$ at fixed temperatures between 125 and 850 mK measured on the powder sample.  }
\label{Xvsf}
\end{figure}

The $M/H$ values obtained in the single crystal are nevertheless far above the powder sample value, as was already seen in the low temperature part of the curves measured in the Quantum Design magnetometer (see Figure 1 of the main text). 
We do not have a clear explanation for these differences. It should be pointed out that in the whole series of pyrochlore iridates, a strong sample dependence of the magnetization vs temperature curves has been observed. It has been ascribed to non magnetic defects, such as Ir$^{5+}$ ions, which alter the molecular field applied on the rare-earth (if it is magnetic) and create pinning centers \cite{suppYang17, suppZhu14}. Even in our samples where the ZFC-FC hysteresis at the metal insulator transition is weak, supporting a weak density of defects, this difference between the susceptibility values suggests that some disorder must be present. 
This is further supported by the magnetization increase observed in the FC curves of the single crystal at about 600~mK, which is not present in the powder samples where the magnetization is almost flat at these temperatures. 

Nevertheless, the single crystal ac susceptibility curve - up to a factor - is similar to the powder sample's one. Especially, the frequency dependence of the peak is the same in both samples (see Figure \ref*{tauvsT}), which shows that the magnetic charge excitations' dynamics is not affected by the differences reported above, and are thus intrinsic to the fragmented phase. 

\section{Cole-Cole analysis}
Measurements of the ac susceptibility as a function of frequency $f$ at fixed temperatures give insight into the dynamics of the system. 
In the presence of a single relaxation time $\tau$, the susceptibility is expected to obey a Debye law 
\begin{equation}
\chi(\omega)= \chi_{\rm S} + \frac{\chi_0 - \chi_{\rm S}}{1+i \omega \tau}
\end{equation}
where $\omega=2 \pi f$, $\chi_0$ and $\chi_{\rm S}$ are the isothermal and adiabatic susceptibilities, respectively. 

It results in a lorentzian shape in the dissipative part $\chi''(f) $ of the susceptibility, centered on a frequency $f_0$ equal to $1/ 2 \pi \tau$, and thus provides a direct determination of the relaxation times of the system. In the Cole-Cole representation, i.e. $\chi"$ vs $\chi'$ plots at a given frequency, the curve is a semi-circle if the Debye law is obeyed, whose radius and center are defined by $\chi_0$ and $\chi_{\rm S}$. 

\begin{figure}[b!]
\includegraphics[width=7.5cm]{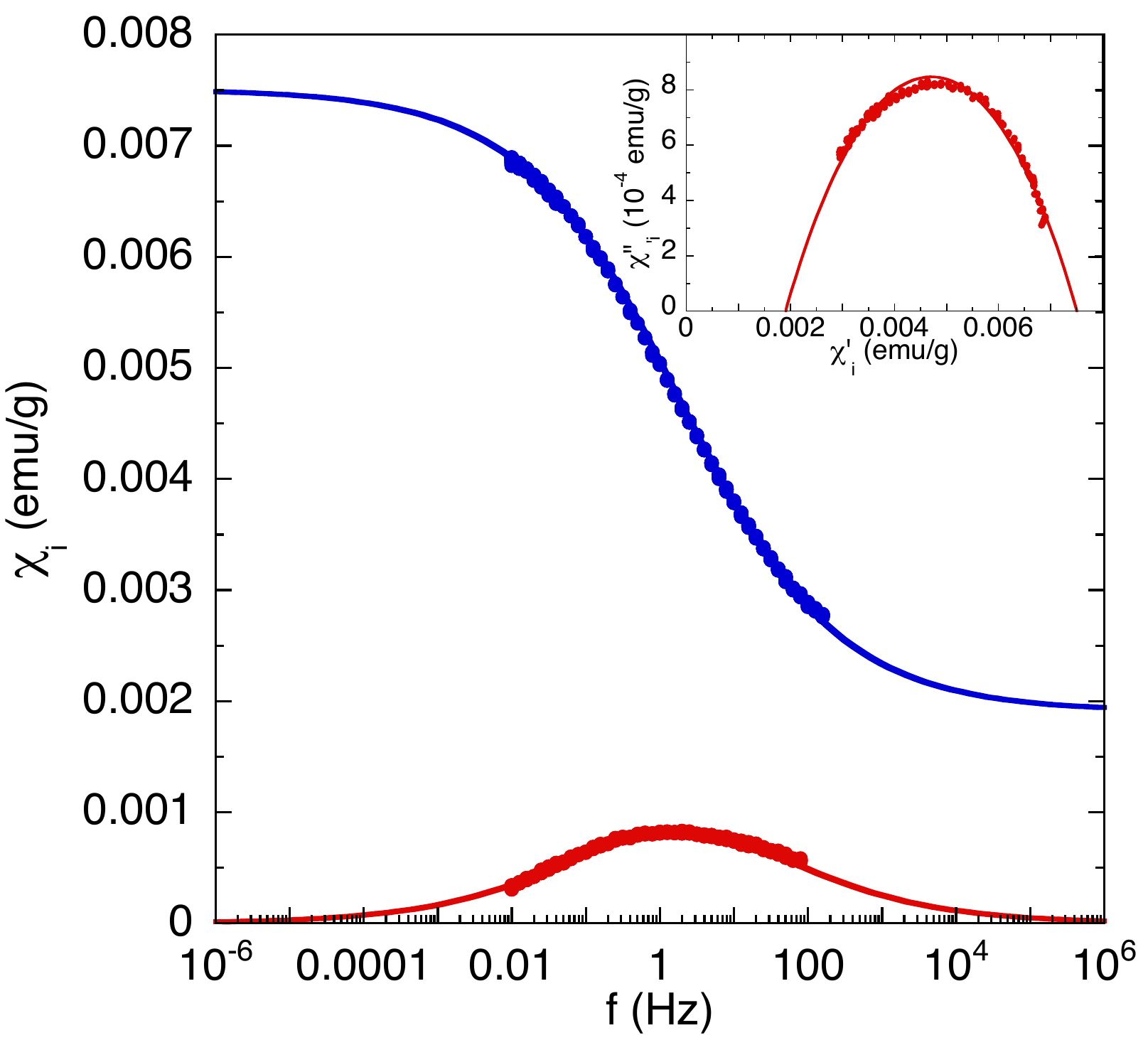}
\caption{$\chi'$ (blue) and $\chi''$ (red) vs $f$ at $T=400$ mK. Inset: $\chi''$ vs $\chi'$. The lines are fits using Equation \ref*{dist_chi} with $\chi_0=0.00751$ emu/g, $\chi_{\rm S}=0.00191$ emu/g, $\tau=0.0865$ s and $\alpha=0.627$. }
\label{fitX}
\end{figure}

In the presence of a distribution of relaxation times centered on a characteristic time, these Cole-Cole plots change into flattened semi-circles. It was shown that the susceptibility can then be described by \cite{suppCole41}:
\begin{equation}
\chi(\omega) = \chi_{\rm S} +  \frac{\chi_0 - \chi_{\rm S}}{1+(i \omega \tau)^{1-\alpha}}
\label{dist_chi}
\end{equation}
where $\alpha$ defines the distribution width. 

In \DIO, the susceptibility $\chi(f)$ curves are much broader than expected from a Debye behavior (see Figure \ref*{Xvsf}). This can be clearly seen in Cole-Cole plots (see inset of Figure \ref*{fitX}), where the curves are strongly flattened semi-circles. At very low temperature, typically below 250~mK, features are so broad that no characteristic time can be defined.  

\begin{figure}
\includegraphics[width=8cm]{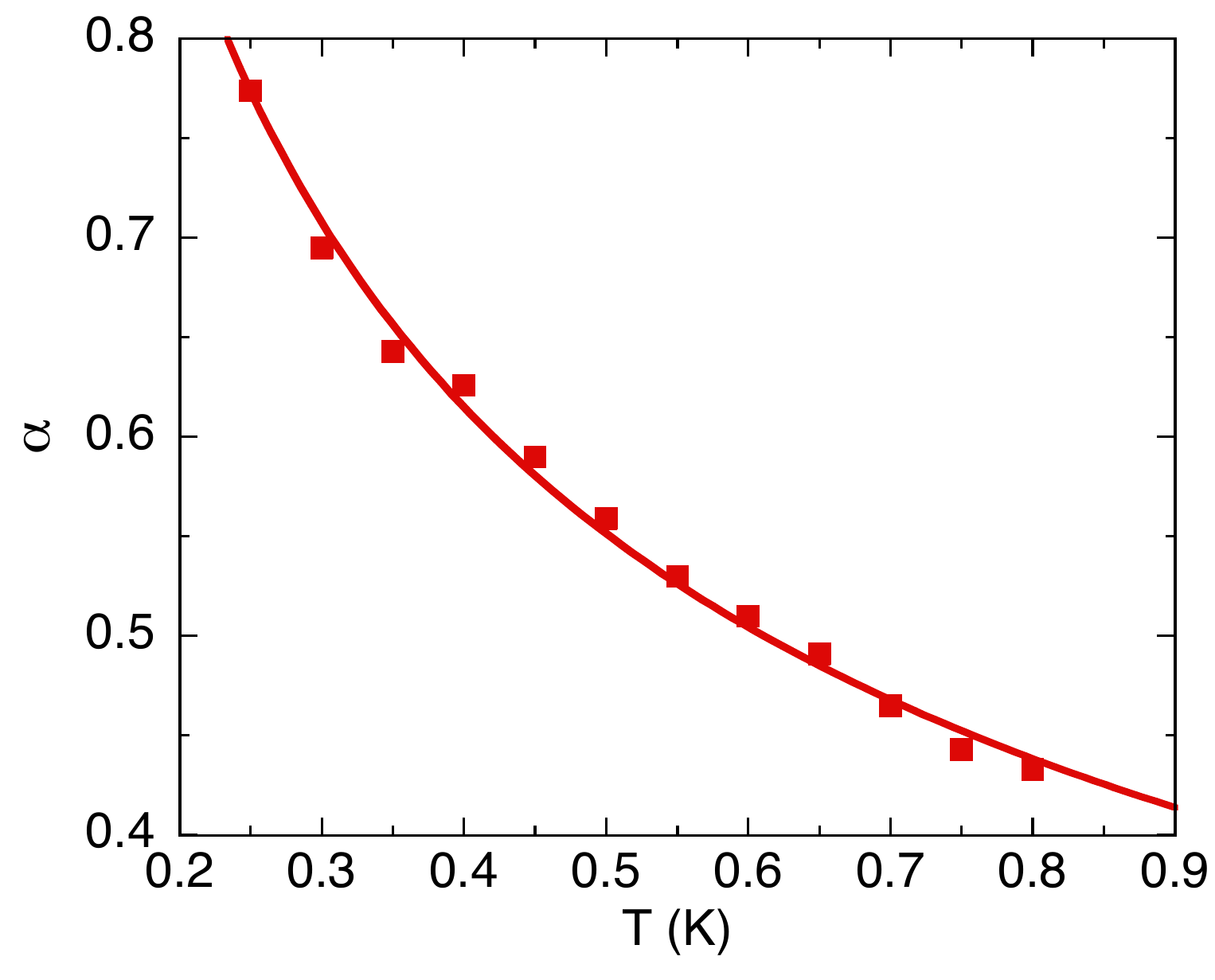}
\caption{$\alpha$ vs $f$, obtained from the fits of the ac susceptibility curves. The line is a fit to the power law: $0.39 \times T^{-0.49}$.}
\label{alpha}
\end{figure}

We have analyzed the susceptibility curves by fitting the frequency dependence of the real part, $\chi'$, and imaginary part, $\chi''$, of the susceptibility, as well as $\chi''$ vs $\chi'$, using the expressions deduced from Equation \ref*{dist_chi} (See Figure \ref*{fitX} for the result at 400 mK) \cite{suppDekker89}.
We then obtain the temperature dependence of the relaxation time shown in the inset of Figure 4 of the main text. In addition, these fits provide an estimation of the distribution width through the $\alpha$ parameter. As expected, it strongly increases when the temperature decreases (see Figure \ref*{alpha}). $\alpha$ seems to follow roughly a $1/T^{1/2}$ dependence whose origin is unknown at the moment.  

\section{Calculations in the dumbbell model}
Data for the dumbbell model were obtained by Monte Carlo simulations with the Metropolis algorithm. 
Long-range interactions were considered by means of Ewald summations \cite{Kaiser14}. 
The internal magnetic fields were modelled by introducing a staggered chemical potential different on A and B sites. 

We simulated cubic systems of $L^3$ conventional unit cells of the pyrochlore lattice, consisting of $16$ spins and $8$ charge sites each, with periodic boundary conditions. 
We used up to $L=8$, and no significant finite-size effects were found. Equilibration took $5\times 10^5$ Monte Carlo steps, and $5\times 10^5$ steps were used to calculate the averaged quantities of interest at each temperature. The results were also averaged over $5$ independent runs.

\section{Excitations in the dumbbell model}

In the monopole picture of spin ice,  the relevant free energy is the grand potential $\Omega=U_C - \mu_1 N_1 - \mu_2 N_2 - ST$ \cite{suppKaiser18} where $U_C$ is the Coulomb energy, $N_1$ and $N_2$ the number of single and double monopoles, $\mu_{1(2)}$ the relevant chemical potentials and $S$ the entropy. 
To calculate the energy cost of an excitation one must consider changes to the ``number enthalpy'', or Landau energy \cite{Landau59} $\delta U=\delta U_C -\mu_1 \delta N_1 - \mu_2 \delta N_2$. The sign convention of thermodynamics is such that $\mu<0$ corresponds to an energy cost for adding a particle. 
In the present problem the notation is further complicated by the staggered chemical potential which favours monopoles of different charge on different sublattices so that north and south monopoles and double monopoles have different chemical potentials for creation on each of them \cite{suppRaban19}.

The lowest energy defect to the monopole crystal turns out to be a double monopole on a site favoured by the staggered term $\Delta$. Introducing this defect goes in two steps: firstly a single monopole is removed from a favoured site. The relevant chemical potential is $\mu_1=\mu+\Delta$ and the change in Landau energy $\delta U= u\alpha+\mu+\Delta$. Secondly one adds a double monopole on a favoured site for which the chemical potential is $\mu_2=4\mu+2\Delta$ with $\delta U= -2u\alpha-4\mu-2\Delta$. The total energy cost is the sum of these two terms, $E_{db}=-(3\mu+\Delta)-u\alpha$, which is the expression given in the main text. Putting in the parameters $u=2.82$ K, $\mu=-4.40$ K, $\Delta=4.95$ K and $\alpha=1.638$ gives $E_{db}=3.63$ K. The energy cost of a monopole hole is the first step of this procedure, $E_{db}^{'}=u\alpha+\mu+\Delta=5.17$~K. 

Which of these two energy scales is largest depends crucially on the relative values of $\mu$ and $\Delta$. This suggests that a set of $\mu$ and $\Delta$ values exist for which $E_{db}=E_{db}^{'}$. The existence of this set of points could have consequences for the stochastic dynamics of the dumbbell model but this point has not been investigated in the present study.

\section{Response of the dumbbell model to a $[111]$ magnetic field}

Taking the magnetic moments as elements of the emergent lattice field, the generation of magnetic monopole quasi-particles leads to the effective fragmentation of the moments into two parts via a Helmholtz decomposition. The first, the ``longitudinal'' part gives the magnetic monopoles and is divergence full. The second, the left over, is divergence free and ``transverse''. The Fourier transforms of the two components are orthogonal to each other. In the monopole crystal phase, the longitudinal part has long range, antiferromagnetic, AIAO order, while the transverse part forms the Coulomb phase classical spin liquid. 

The applied field acts on both the longitudinal and transverse parts. The field provides a potential energy gradient for the monopoles so that the north and south poles reduce their energy by moving in opposite directions with respect to it. However, the constraints of magnetism exclude the possibility of a {\it dc} monopole current \cite{suppRyzhkin05,suppJaubert09}. With the field placed along the $[111]$ direction the combination of energy gradient and constraints produces an effective staggered chemical potential for the monopoles \cite{suppRaban19}. However, in this supplementary information we simply treat the Zeeman energy of the spins in addition to the Coulomb energy of the monopoles, leaving a complete discussion of the effect of the field on the two components for future work.  

The four spins of a unit cell, taken here to consist of an ``up'' tetrahedron, lie parallel or anti-parallel to the four body centred cubic axes 
\begin{equation*}
\begin{aligned}
&{\bf d_1}=\frac{1}{\sqrt{3}}[1,1,1], & {\bf d_2}=\frac{1}{\sqrt{3}}[1,-1,-1],\\
&{\bf d_3}=\frac{1}{\sqrt{3}}[-1,1,-1], & {\bf d_4}=\frac{1}{\sqrt{3}}[-1,-1,1].
\end{aligned}
\end{equation*}

The ice rules with, for each tetrahedron two spins in and two out, correspond to two spins aligned and two anti-aligned. A monopole with charge $Q$ ($-Q$) sits on a tetrahedron with ``three-in/one-out'' (``three-out/one-in'') and corresponds to three spins aligned (anti-aligned) and one anti-aligned (aligned). Taking the convention that the staggered internal field of the iridium ions favours a positively charged monopole on an up tetrahedron, the forward (reverse) magnetic field is ${\bf H} = +(-) \frac{H}{\sqrt{3}}[1,1,1]$.

\begin{figure}[h!]
\centering
\includegraphics[width=7.5cm]{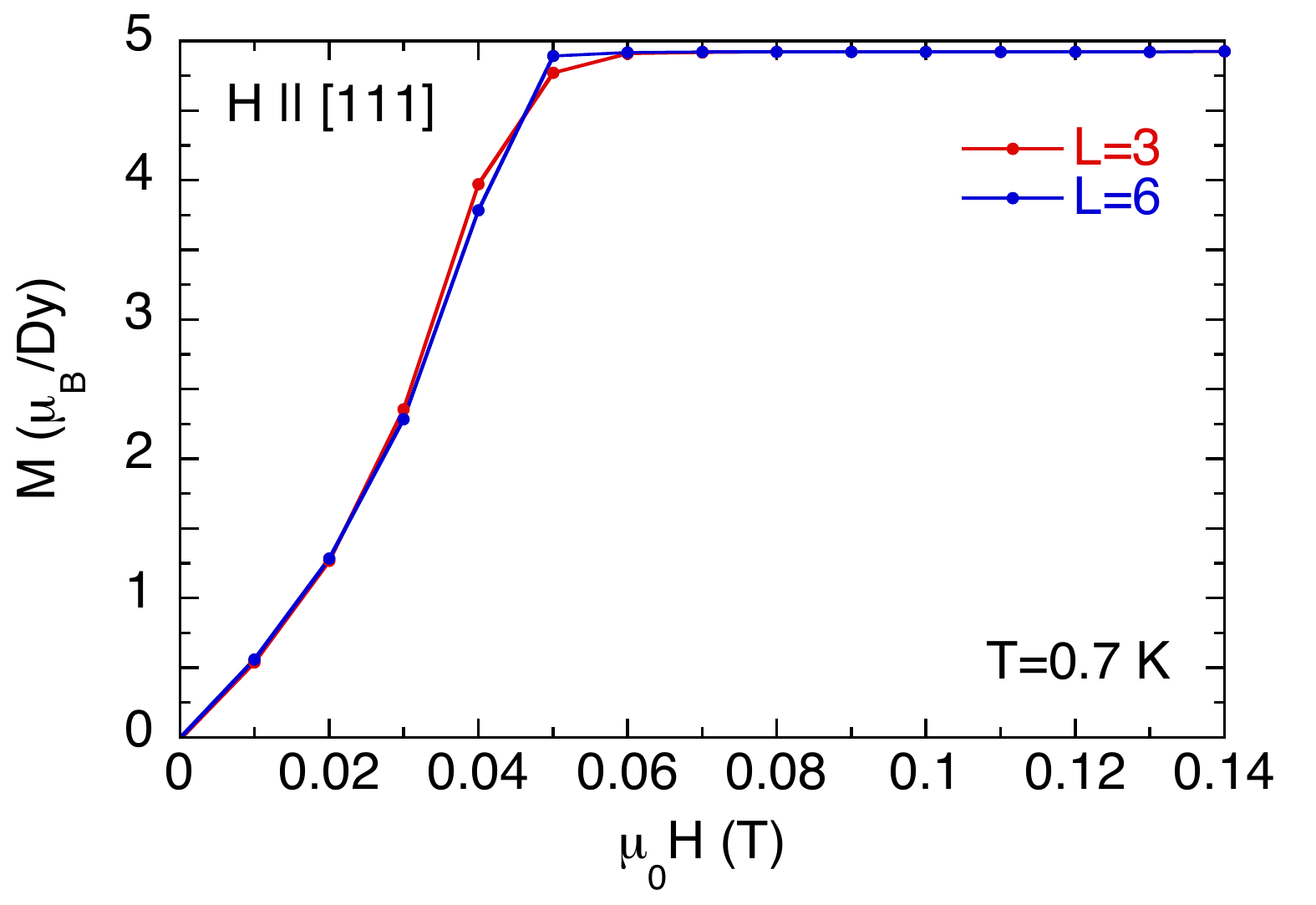}
\caption{\label{f-field} Calculated $M$ vs $H$ with the field in the forward direction, ${\bf H} \parallel [111]$, for $T=0.7$ K and two system sizes ($L=3$ and 6).}
\end{figure}

With the field in the forward direction, there is no change in the monopole ordering on increasing the field. Above the saturation field the apical spin on the up tetrahedron points out along the field direction with the three spins of the base pointing in, each with projection of $m/3$ along the field direction giving a total projection per spin along the field direction of $m/2$. 
Ferromagnetic order comes from extending the loops of zero energy spins flips (the loops of hard core dimers) into system spanning topological sectors \cite{Jaubert13}. Starting from the saturated ordered state, excluding any defects in the ionic crystal the system can only disorder via a Kasteleyn transition \cite{suppMoessner03,suppBrooks14}. To see this one must consider the construction of a loop of reversed spins passing through a unit cell of the pyrochlore lattice. 

The flipped loop enters the tetrahedron by flipping the spin at its apex and leaves by flipping one of the three spins on the base. The total change in Zeeman energy for these combined flips is $\delta\epsilon=\dfrac{8\mu_0mH}{3}$ while the generated entropy is $\delta s=k_{\rm B}\ln{(3)}$ giving a contribution to the free energy, $\delta \Omega=\delta \epsilon-T\delta s$. Placing the loop therefore reduces the free energy for $T>T_K=\dfrac{8\mu_0mH}{3k_{\rm B}\ln{(3)}}$ where $T_K$ is the Kasteleyn transition temperature. Putting our numerical value of $m=9.85~\mu_{\rm B}$ we find a universal ratio at the Kasteleyn transition: 
\begin{equation} 
\frac{\mu_0 H}{T}=0.062 \;\;{\rm T.K}^{-1}
\end{equation}

In Figure \ref*{f-field} we show preliminary simulation data for magnetisation as a function field for the dumbbell model applied in the forward direction, at $T=0.7$ K for $L=3$ and $L=6$. Our predicted critical field at this temperature is $\mu_0 H=0.0435$ T. The data show a sharpening with system size towards a singularity in the saturated magnetisation at a field close to the predicted value and the data is consistent with a Kasteleyn transition \cite{suppMoessner03}. We anticipate that more sophisticated simulations using a non-local loop algorithm \cite{Jaubert08} would confirm this prediction with precision. At the temperature used in the experiment, $T=0.1$ K, the critical field would be $\mu_0H=6.2$~mT. 
\bigskip

For the field in the reverse direction, the magnetisation saturates at a first plateau value for small field, with the apical spin pointing in, along the field direction. Two of the base spins point in, with projection against the field with one pointing out, projecting with the field (forward crystal). The projected moment per spin is thus $m/6$. Increasing the field further reduces the effective value of $\Delta$ until a threshold field passes the system back into the spin ice state with two spins in and two out. That is, one of the basal spins flips giving a second plateau with projection of $m/3$ per spin. Increasing the field further, the final basal spin flips above a second threshold giving a three-out/one-in tetrahedron, that is a south pole on a site where the internal field on its own would favour a north pole (reversed crystal). 

The field thresholds can be estimated at zero temperature by calculating the minimum Landau energy for the three phases. From equation (2) - main text:
\begin{equation}
\begin{aligned}
U_1&=N_0\left[-\frac{u\alpha}{2} - \mu - \Delta - \frac{\mu_0 m H}{3}\right], \\
U_2&=N_0\left[ - \frac{2\mu_0 m H}{3}\right],\\
U_3&=N_0\left[-\frac{u\alpha}{2} - \mu + \Delta - \frac{\mu_0 m H}{3}\right], 
\end{aligned}
\end{equation}
where $U_1$ corresponds to the forward monopole crystal phase, $U_2$ the spin ice phase and $U_3$ the reversed monopole crystal phase.

The field thresholds correspond to $U_1=U_2$ and $U_2=U_3$, coexistence between phases $1$ and $3$ being thermodynamically unstable. It follows that
\begin{equation}
\begin{aligned}
\mu_0 H_1&=&\left(\frac{3k_B}{m}\right)\left[\frac{u\alpha}{2} + \mu + \Delta\right], \\
\mu_0 H_2&=&\left(\frac{3k_B}{m}\right)\left[-\frac{u\alpha}{2} - \mu + \Delta\right].
\end{aligned}
\end{equation}
Using parameters from the main text, $u=2.82$ K, $\mu=-4.4$ K, $\Delta=4.95$ K and $\alpha=1.638$ we find $\mu_0 H_1=1.3$~T and $\mu_0H_2=3.2$~T respectively.

\begin{figure}[h!]
\centering
\includegraphics[width=7.5cm]{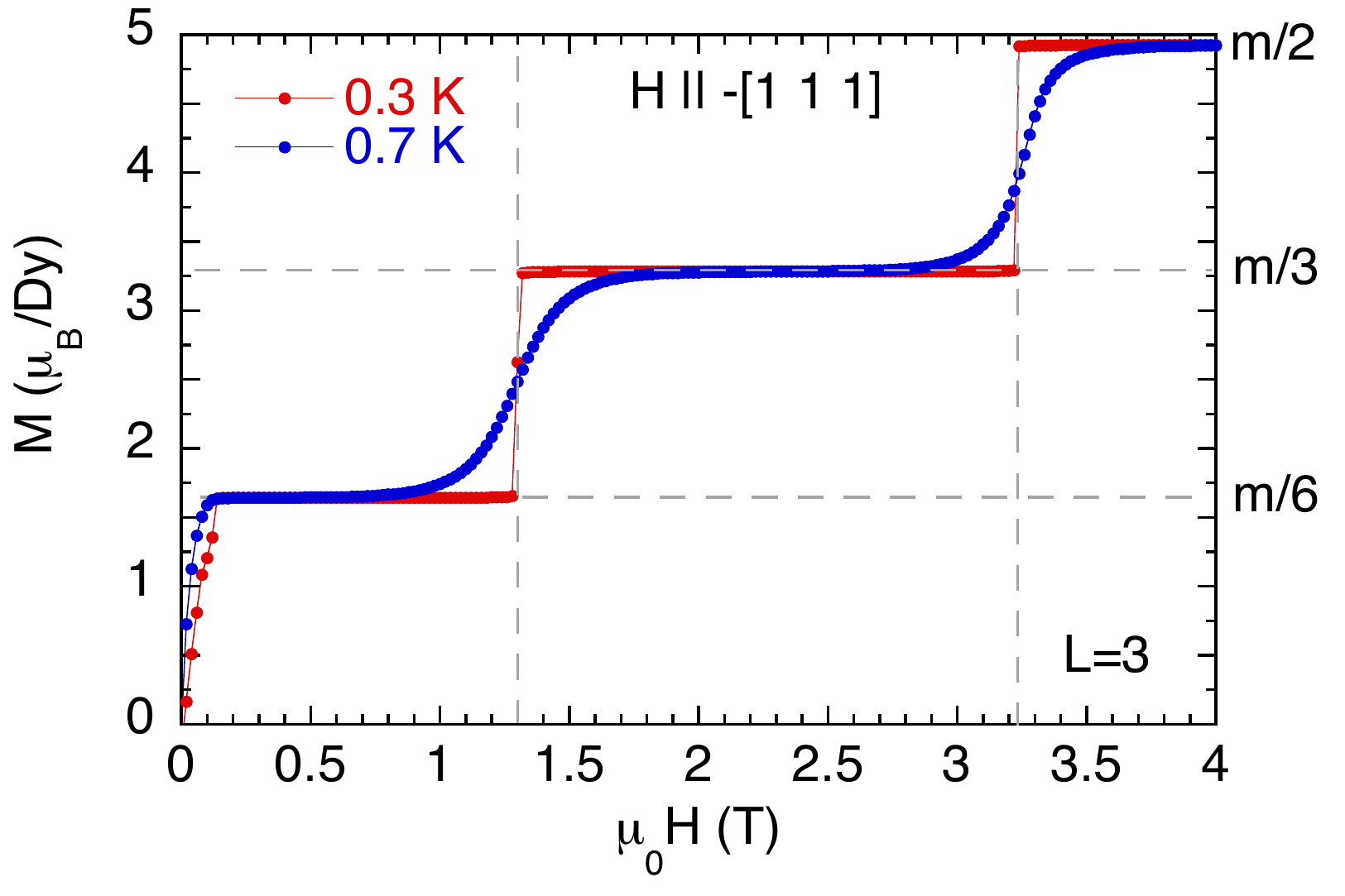}
\caption{\label{r-field} Calculated $M$ vs $H$ with the field in the reverse direction, ${\bf H} \parallel [$-1-1-1$]$, for $T=0.3$ and 0.7 K for $L=3$.}
\end{figure}

In Figure \ref*{r-field} we show magnetisation against field simulated from the dumbbell model for $L=3$ for $T=0.3$~K and $0.7$ K. The data are seen to follow our predictions accurately and after a rapid rise, magnetisation plateaus are observed at the predicted values for the predicted fields. Note that, while at $T=0.3$ K the evolution between the plateaus is discontinuous, at $T=0.7$ K the jumps become rounded. This strongly suggests the existence of phase transitions in the family of transitions outlined in Ref. \onlinecite{suppRaban19}, in which case one would expect the transitions to end at a critical end point somewhere above $T=0.35$ K but this point was not pursued in the present study.


\end{document}